\documentclass[preprint2]{aastex6}

\begin{document}

\title[Discovery of tidal tails around NGC~7492]{The discovery of
  tidal tails around the globular cluster NGC~7492 with Pan-STARRS1}

%% Use \author, \affil, plus the \and command to format author and affiliation 
%% information.  If done correctly the peer review system will be able to
%% automatically put the author and affiliation information from the manuscript
%% and save the corresponding author the trouble of entering it by hand.
%%
%% The \affil should be used to document primary affiliations and the
%% \altaffil should be used for secondary affiliations, titles, or email.

%% Authors with the same affiliation can be grouped in a single
%% \author and \affil call.

\author{
C. Navarrete\altaffilmark{1,2,3}, 
V. Belokurov\altaffilmark{3}, and 
S. E. Koposov\altaffilmark{3,4}}
% List of institutions
\altaffiltext{1}{Instituto de Astrof\'isica, Pontificia Universidad Cat\'olica de Chile,  Av. Vicu\~na Mackenna 4860, 782-0436 Macul, Santiago, Chile.}\email{cnavarre@astro.puc.cl}
\altaffiltext{2}{Millennium Institute of Astrophysics, Santiago, Chile.}
\altaffiltext{3}{Institute of Astronomy, University of Cambridge, Madingley Road, Cambridge, CB3 0HA, UK.}
\altaffiltext{4}{McWilliams Center for Cosmology, Department of Physics, Carnegie Mellon University, 5000 Forbes Avenue, Pittsburgh, PA 15213, USA.}

\begin{abstract}
 
We report the discovery of tidal tails around the Galactic globular cluster 
NGC~7492, based on the Data Release 1 of the Pan-STARRS 1 survey. The tails were 
detected with a version of the matched filter technique applied to the $(g-r,r)$ 
and $(g-i,i)$ color-magnitude diagrams. Tidal tails emerging from the cluster 
extend at least $\sim$3.5$\degr$ in the North-East to South-East direction, equivalent to 
$\sim1.5$ kpc in projected length.

\end{abstract}


\section{Introduction} \label{sec:intro}

Globular clusters with tidal tails are precious for three reasons. First, there 
exists a coupling between the strength of the
tides experienced by the cluster and its internal dynamics \citep[see 
e.g.][]{Gnedin99}. Therefore, through simultaneous modelling of the bound and 
unbound stellar mass distributions, one can learn the details of the satellite's 
evolution in the Galaxy \citep[see e.g.][]{Dehnen04}. Second, stellar tails grow 
approximately along the cluster's orbit \cite[e.g.][]{Eyre11}, thus providing a 
powerful technique to infer the properties of the host potential itself 
\citep[e.g.][]{Koposov10,Bowden15,Kupper15}. Third, globular cluster tails are 
fragile enough to be easily perturbed by low-mass objects, for example, dark 
matter subhalos with masses below $10^8$M$_{\odot}$ 
\citep[e.g.][]{Yoon11,Erkal2015b}. Thus globular cluster streams are a unique 
tool for measuring the lumpiness of the Galactic dark matter distribution 
\citep[][]{Erkal15a,Erkal16}.

A large number of the Milky Way globulars are predicted to be undergoing 
destruction today \citep[see][]{Gnedin97}, however so far, tails have been 
detected only around a handful of objects \citep[e.g.][]{Odenkirchen2001, 
Belokurov2006, NO10, Solima11, Lauchner2006}. This discrepancy could perhaps be 
alleviated by taking into account the effects of mass segregation: the stars 
shed by the cluster in the initial phases of the dissolution are simply too 
low-mass to light up the tails \citep[see][]{Balbinot17}. Additionally, this 
hypothesis may help to explain a substantial number of orphan stellar streams 
revealed so far
\citep[see e.g][]{Grillmair06, Bonaca12,Koposov14,Balbinot16}. Most importantly, 
it serves to re-invigorate the search for tidal tails around globular clusters, 
albeit at lower surface brightness levels.

The absolute majority of the discoveries mentioned above have been made using 
the data from the Sloan Digital Sky Survey (SDSS), which by now has been trawled 
extensively for the Galactic stellar halo sub-structure. Motivated by the 
prospects of unearthing new examples of tidal disruption, we have searched for 
the presence of stellar tails around globular clusters located within the 
footprint of the Pan-STARRS1 (PS1) 3{$\pi$} survey. Recently, after years of 
anticipation, the PS1 object catalogues were finally released publicly and are 
currently accessible for download through the MAST archive 
\citep[see][]{Flewelling16}. Compared to the SDSS, PS1 i) provides continuous 
coverage at low Galactic latitudes in the Northern hemisphere and ii) extends 
$\sim$30$^{\circ}$ further down under the celestial equator. Note that ours is 
not the first exploration of the
Galactic stellar halo sub-structure with PS1 \citep[see e.g.][]{Bernard2014, 
Bernard2016}.

\begin{figure*}
\includegraphics[width=0.3\textwidth]{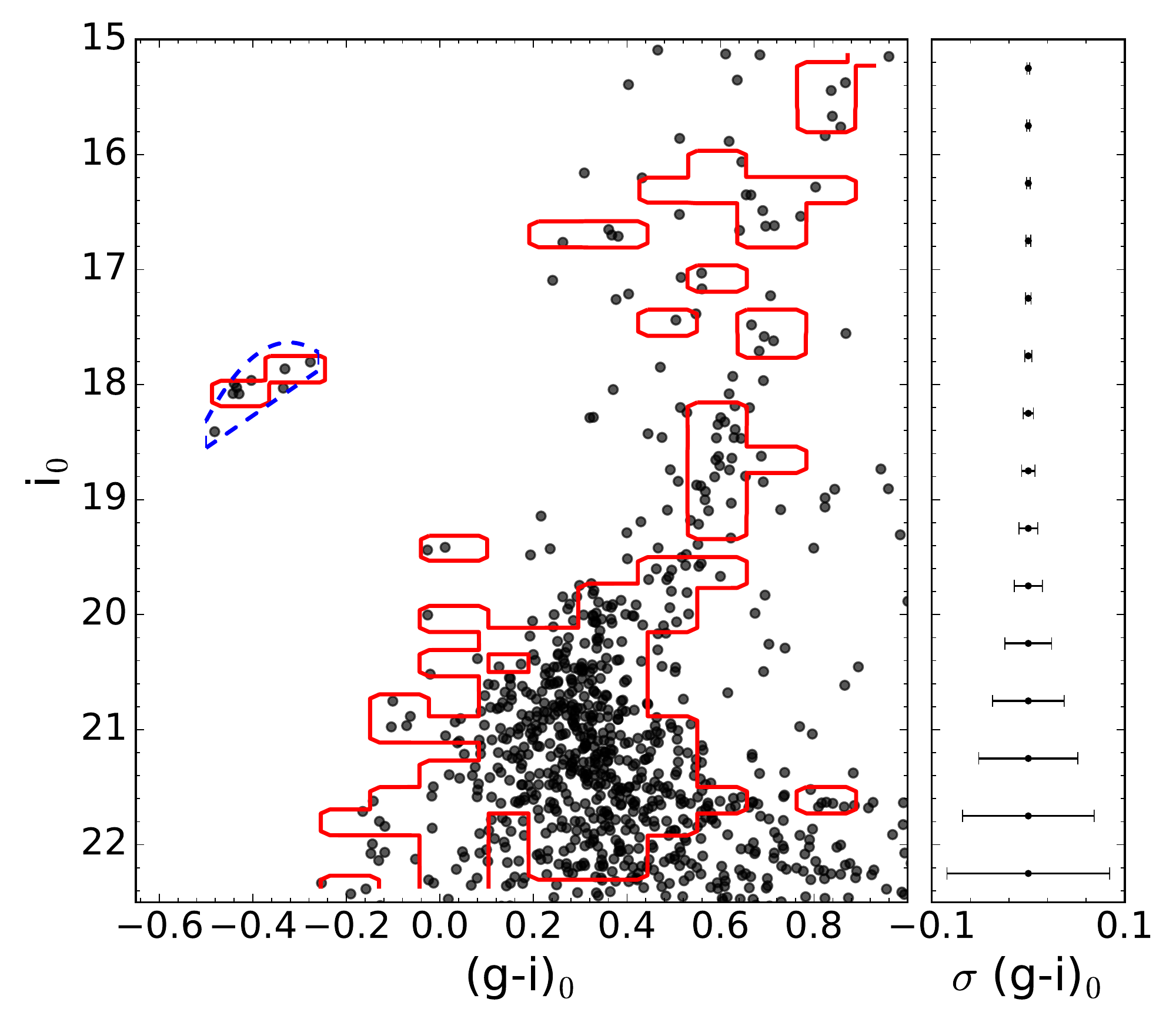}
\includegraphics[width=0.6\textwidth]{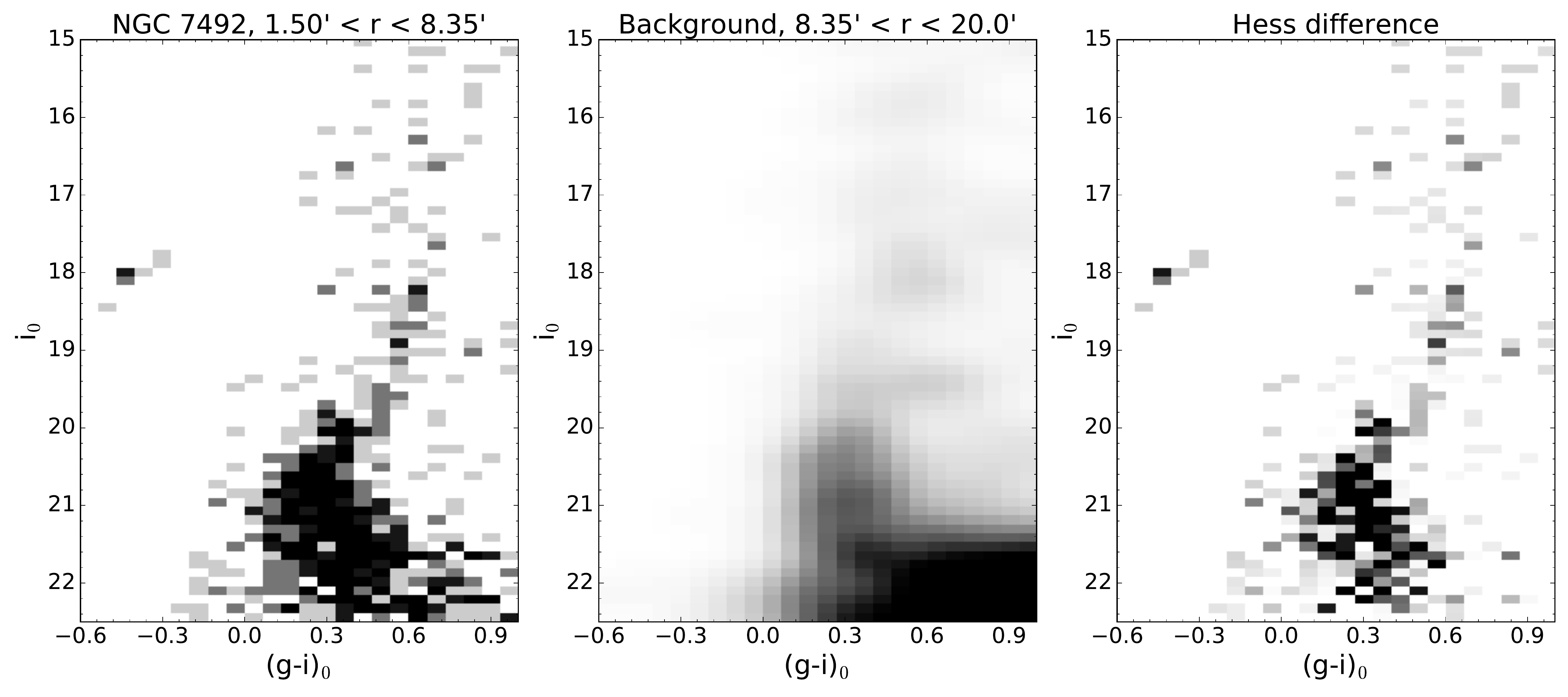}
\caption{{\it First panel:} Extinction-corrected $(g-i, i)$ color-magnitude 
diagram of NGC~7492. It only includes stars within the tidal radius of the 
cluster ($r_t = 8.35\arcmin$), excluding the innermost center. The blue polygon 
encloses the BHB population of the cluster. The red contours show the 
color-magnitude bins where the matched filter selected stars in a 
4$\times$4{\degr} field around the cluster fall into. The right inset shows the 
mean $(g-i)$ color photometric error as a function of $i$ magnitude. Note that 
the apparent scatter around the cluster's CMD is larger than the photometric 
errors. Some of the contamination must come from the Galactic 
foreground/background (shown in the third panel). {\it Second panel:} Hess 
diagram for stars inside the tidal radius of the cluster, but excluding the 
innermost 1.5$\arcmin$. {\it Third panel:} Hess diagram for stars with 
8.35$\arcmin < r < 20.0\arcmin$ away from the cluster's center. Sgr stream could 
provide some contribution to the MS and MSTO region of this CMD. {\it Fourth 
panel:} Hess difference. A number of familiar stellar populations can be 
identified in the second and fourth panels: MS, TO and HB as well as sparsely 
populated blue straggler sequence. In the background subtracted Hess diagram 
most of the contamination is removed and the sequences are much narrower.
\label{fig:hess}}
\end{figure*}

In this Letter, we report the discovery of tidal tails around NGC~7492, a sparse 
outer halo Galactic globular cluster (R$_{\rm GC} \sim$ 25 kpc; 
\citealt{Harris96}, 2010 edition; see also \citealt{Figuera13}). Previous 
studies of the cluster suggested the presence of extra-tidal stellar material. 
For example, \cite{Leon00} found a small extension pointing towards the Galactic 
center, but emphasized the need for high-quality deep CCD data to confirm this 
low surface-brightness structure unambiguously. Later, studying a field 
$42\arcmin \times 42\arcmin$ in size, \cite{Lee04} noticed tail-like structures 
extending towards the northeast and northwest from the cluster. Given its mass 
and Galacto-centric distance, NGC~7492 was also tagged as a ``tidally-affected'' 
cluster by \cite{Julio12}. Other clusters included in this category are already 
known to have tidal tails, such as Pal~5 \citep{Rockosi2002, Odenkirchen2003, 
Grillmair2006}, NGC~5466 \citep{Belokurov2006}, and NGC~5053 
\citep{Lauchner2006}.

\section{PS1 data} \label{sec:ps1_data}

The Panoramic Survey Telescope and Rapid Response System (Pan-STARRS, PS1) 
survey is observing the sky north of declination $-30\degr$ with a 1.8m optical 
telescope from the summit of Haleakala, in Hawaii. The telescope has a 7 deg$^2$ 
field of view, imaged with a mosaic CCD camera with 60 detectors, each with 4800 
$\times$ 4800 pixels. The pixel scale is 0.258 arcsec. The images are obtained 
through five filters: $gP1$, $rP1$, $iP1$, $zP1$ and $yP1$. The first data 
release (DR1) contains object catalogues produced using both non-stacked and 
stacked catalogues of the 3$\pi$ Steradian Survey. Details of the data delivered 
in PS1 DR1 are described in detail by \cite{Chambers16, Magnier2016a, 
Magnier2016b}. In the analysis presented here, stars were selected from the PS1 
stack detection catalogue considering only sources with {\tt (rpsfmag-rkronmag)} 
$< 0.05$ \citep[i.e. most likely stellar objects, see][]{Farrow2014} and $g$ and 
$r$ {\tt STACK\_PRIMARY} flag-set greater than zero. The faintest magnitude 
considered was $r =$~22.5 mag. The apparent magnitudes were corrected for 
extinction using the dust maps of \cite{SFD} as calibrated by \cite{Schlafly11}.

The left panel of Figure~\ref{fig:hess} shows the Color-Magnitude diagram
(CMD) of the stars located inside the tidal radius of the cluster
\citep[$r_t \sim 8.3\arcmin$,][]{Harris96}. To remove stars suffering
from crowding, only objects outside the very center (i.e, $r >
1.5\arcmin$) are included. Despite contamination from field stars, the
Main Sequence (MS), sub-giant branch, Red Giant Branch (RGB) and Blue
Horizontal Branch (BHB) are clearly distinguishable. The CMD of the
cluster also reveals the presence of a small number of stars brighter
than the MS turn-off (TO), most probably blue straggler stars, as
previously claimed by \cite{Cote91}. The RGB appears somewhat
broadened: this scatter is difficult to explain given the current
constraints on chemical abundance variations in the cluster, but very
few giants from the cluster have spectroscopic studies \citep[see][and
  references therein]{Cohen2005}.

The three rightmost panels of Figure~\ref{fig:hess} show the Hess diagrams
(stellar density in color-magnitude space) for a field centered on the
cluster (middle left panel), for an outer annulus (middle right), and
their difference (right panel). The Hess diagram for the cluster field
gives a clearer view of the satellite's stellar populations: an
extended BHB at $-0.6 < (g-i) < -0.3$ and $i \sim 18$ mag; the MS
populated mainly by stars fainter than $i = 20$ mag; and the TO point
at $(g-i) = 0.4$ and $i \sim 20$ mag. While overdensities of stars
around these evolution stages are clearly discernible, there is
substantial contamination at the faintest and reddest magnitudes, as
well as in the RGB region. The decontaminated Hess difference (right
panel of the Figure) allows to peel away the foreground layer and thus
makes the cluster's stellar sequence much more evident. Here,
reassuringly, the satellite's MS appears much tidier and even hints of
the asymptotic-giant branch are noticeable at magnitudes brighter than
$i \sim 17$ mag.

Note that the cluster location on the sky at ($\alpha$, $\delta$) =
($347.1{\degr}$, $-15.6{\degr}$) (J2000), or $l = 53\fdg 38$, $b = -63
\fdg 48$, NGC~7492 lies within the projected position of the
Sagittarius (Sgr) trailing stream as pointed out by
\citet{Julio14}. Curiously, among all Galactic globular clusters in
the vicinity of the stream, it has the lowest probability to be
associated with the dwarf according to the model of
\cite{LM10}. Nonetheless, we conjecture that a good fraction of the
field contamination around the NGC~7492 is supplied by the Sgr debris,
given that both systems are at similar heliocentric distances and
located relatively close on the sky \citep[see Figure 17
  from][]{Julio14}.
  
\section{Detection of the tidal tails  and discussion} \label{sec:results}

To search for extra-tidal extensions to the cluster's light
distribution, we broadly follow the matched filter methodology as
described in \cite{Rockosi2002}. Taking advantage of the PS1
multi-band photometry we generate matched filter masks for both
$(g-r,r)$, and $(g-i,i)$ CMDs. Only stars with $ 15.0 < i, r < 22.5$
were considered. These magnitude cuts avoid possible saturated stars
and the faint stars with larger photometric errors. The
color-magnitude distribution of the matched filter selected stars was
constructed using the background-subtracted Hess diagrams for
$(g-i,i)$ (rightmost panel of Figure~\ref{fig:hess}) and the
equivalent $(g-r,r)$ Hess subtracted diagram. For the
foreground/background color-magnitude distribution, an adjacent area
of the sky was used, excluding a window in right ascension around
NGC~7492. The background CMD was constructed from four regions,
defined at the boundaries: $339\degr <$ R.A. $< 342\degr$, $342\degr
<$ R.A. $< 345\degr$, $349\fdg25 <$~R.A.~$< 352\fdg25$, $352\fdg25 <$
R.A. $< 355\fdg25$, and between $-20\degr <$ Dec. $< -10\degr$. The
Hess diagrams of the four fields were averaged to obtain a mean
background density of stars in the color-magnitude plane.

\begin{figure}
\centering
\includegraphics[width=0.45\textwidth]{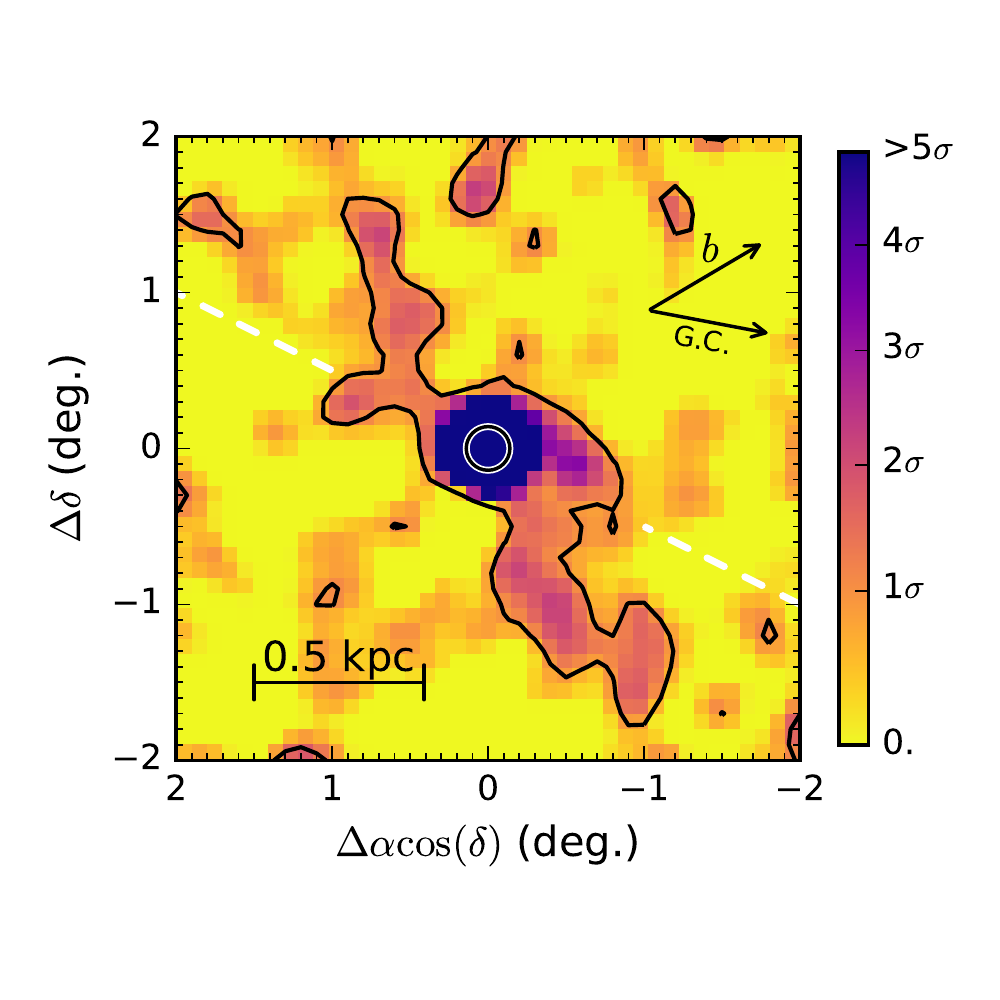}
\caption{Logarithm of the density of stars that passed the two matched filters (see 
main text). The map is 43$\times$43 pixels across and was smoothed with a 
Gaussian filter with a standard deviation of 0.12${\degr}$. The scale between 
degrees and the physical distance, in kpc, is indicated in the bar at the left 
bottom corner. The innermost circle corresponds to the tidal radius of the 
cluster. The contour encloses the pixels with weights that are $>1 \sigma$ above 
the mean background. Note two narrow stellar tails extending symmetrically from 
the cluster. As a reference, in the upper right corner, the direction towards 
the Galactic center (G.C.) and towards increasing Galactic latitude, $b$, values 
are indicated by the arrows. The dashed white line marks the direction of the 
Sgr stream stars, according to the model from 
\cite{LM10}.\label{fig:tidaltails}}
\end{figure}

The ratio of the cluster's and the Galactic background's CMD densities (with the 
pixel size of 0.12$\times$0.2 mag) was used to define the selection region in 
the $(g-r,r)$ and $(g-i, i)$ planes independently. Any star can then be assigned 
a weight corresponding to the value of the Hess ratio of the CMD pixel it falls 
into. For instance, BHB stars tend to have higher weights since the field 
color-magnitude distribution is less populated in that region. The weight 
assigned was the sum of the weights in the $(g-r,r)$ and $(g-i,i)$ planes. Using 
these two matched filters (or three photometric bands) simultaneously improves 
the signal of the stream significantly. It helps to reduce the impact of small 
fluctuations in the CMD of the cluster, which is sparsely populated to begin 
with, and additionally suffers from contamination from field stars even in the 
cluster innermost area (see Figure~\ref{fig:hess}). Once the weight was assigned, 
only stars with weights above a certain threshold were considered as most likely 
cluster members. The weight threshold is determined by maximising the 
signal-to-noise of the cluster
itself. This is the same approach as used by \cite{Erkal16}. The red contour in 
the left panel of Figure~\ref{fig:hess} shows the CMD bins where the matched 
filter selected stars, from Figure~\ref{fig:tidaltails}, fall into. Despite some 
possible contamination from stars belonging to the Sgr stream, the stars 
selected seems to follow the MS and RGB of the cluster.

Figure~\ref{fig:tidaltails} shows the density of the matched filter selected 
stars in the 4$^{\circ}$x4$^{\circ}$ (1.8 $\times$ 1.8 kpc) area centered 
on the cluster. The map was smoothed using a Gaussian filter with $\sigma = 
0\fdg12$. The innermost circle marks the tidal radius of the cluster and the 
contour line confines the pixels with values at 1 $\sigma$ above the background. 
As evidenced from the Figure, the matched filter reveals two tails on either 
side of the cluster, extending over $\sim 3.5 \degr$ ($\sim 1.5$ kpc length) in 
the North-South direction. This structure resembles the characteristic 
``S-shaped'' tidal feature found around other disrupting globular clusters, such 
as Pal 5 and NGC~5466. The contours drawn by \cite{Lee04} are consistent with 
the northern branch and the small lobe found at $(\Delta \alpha \cos{\delta}, 
\Delta \delta) \sim (-0\fdg7, -0\fdg1)$. Our results extend the northern branch 
on a factor of $\sim5$ in length and also recover the southern branch, which was 
not seen previously. Given that there are no previous measurements of 
proper motions for this cluster, nor measurements in the Tycho–Gaia Astrometric 
Solution \citep[TGAS][]{Michalik2015, Lindegren2016} catalogue, it is not 
possible to compare the direction of the stream with the projection of the orbit 
of the cluster, but in principle, both should be aligned. The white dashed line 
in Fig.~\ref{fig:tidaltails} marks the direction of the Sgr stream stars, 
according to the model of \cite{LM10}. Despite the cluster is projected onto the 
expected location of the Sgr tidal stream, the dwarf's leading arm's direction 
is misaligned with the tidal tails detected in this work, thus making the 
possibility of an association the two rather tenuous.

\begin{figure}
\centering
\includegraphics[width=0.48\textwidth]{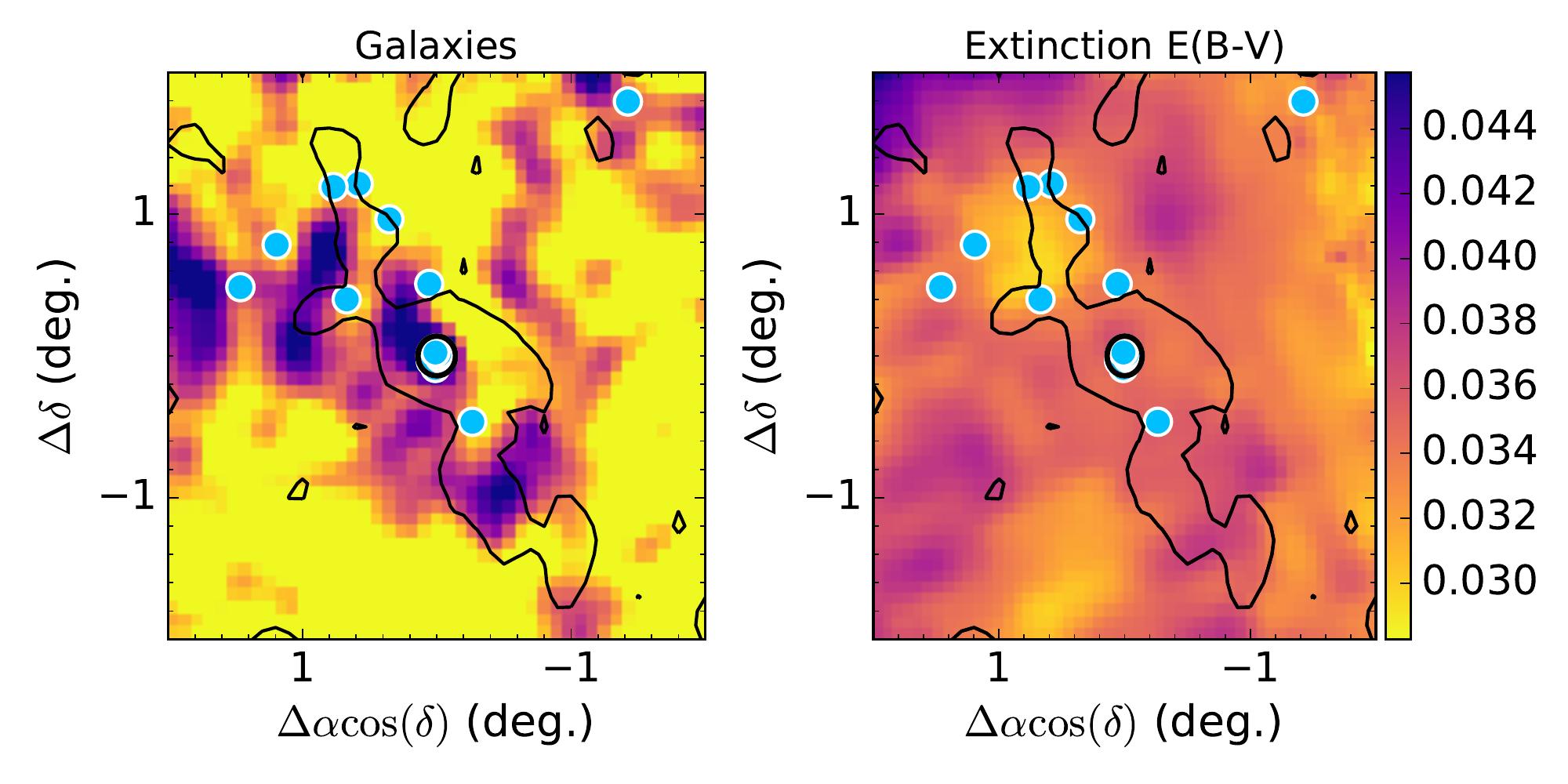}
\caption{{\it Left:} Number density of galaxies around the NGC~7492 globular 
cluster. The bin size and smoothing length are the same as those used in 
Figure~\ref{fig:tidaltails}. The innermost circle corresponds to the tidal 
radius while the contour line to the pixels above 1$\sigma$ in the matched 
filter map of stars around the cluster (defined in Figure~\ref{fig:tidaltails}). 
Light blue filled circles give the positions of the BHB candidate stars, at the 
distance of NGC~7492. {\it Right:} $E(B-V)$ extinction around the cluster. The 
bin size and lines are the same as in the left panel. The colour bar shows the 
mean extinction value per pixel. \label{fig:galaxies_extinction} }
\end{figure}

As a final check, we investigate whether some of the features seen in 
Figure~\ref{fig:tidaltails} could be caused by misclassified galaxies and/or 
effects of the interstellar extinction. The left panel of 
Figure~\ref{fig:galaxies_extinction} shows the logarithmic spatial distribution 
of galaxies (i.e.  sources with {\tt (rpsfmag-rkronmag)} $\geq 0.05$) around 
NGC~7492, selected based on the same matching filters as in the stellar density 
map and using the same bin size and smoothing kernel as in 
Figure~\ref{fig:tidaltails}. To guide the eye, the tidal radius of the cluster 
and the 1$\sigma$ contour of the matched filter density are over-plotted. There 
seems to be no direct correlation between the tails and the distribution of 
galaxies in the field. The variations in the extinction are unlikely to have 
affected our results as indicated in the right panel of 
Figure~\ref{fig:galaxies_extinction}. Clearly, the overall reddening of the 
Galaxy in the area is rather low (ranging between 0.03 and 0.045 mag) and its 
variation across the field does not correlate with tidal tails detected. 
Figure~\ref{fig:galaxies_extinction} also gives the positions of the BHB 
candidate stars, shown as light blue circles. These were selected to lie within 
the blue polygon in the left panel of Figure~\ref{fig:hess}. Despite their 
overall low number density, BHBs appear to follow the Northern tail, connecting 
its uppermost tip with NGC~7492 itself. On the other side of the cluster, 
however, there is a distinct lack of the BHB candidates. This perhaps is not 
surprising given the low number of BHB stars expected in a given stellar 
population. While the spatial distribution of the possible BHB candidates is 
suggestive, deeper wide imaging is required to confirm (and perhaps extend) the 
discovery presented here.

\acknowledgments
The Pan-STARRS1 Surveys (PS1) and the PS1 public science archive have
been made possible through contributions by the Institute for
Astronomy, the University of Hawaii, the Pan-STARRS Project Office,
the Max-Planck Society and its participating institutes, the Max
Planck Institute for Astronomy, Heidelberg and the Max Planck
Institute for Extraterrestrial Physics, Garching, The Johns Hopkins
University, Durham University, the University of Edinburgh, the
Queen's University Belfast, the Harvard Smithsonian Center for
Astrophysics, the Las Cumbres Observatory Global Telescope Network
Incorporated, the National Central University of Taiwan, the Space
Telescope Science Institute, the National Aeronautics and Space
Administration under Grant No. NNX08AR22G issued through the Planetary
Science Division of the NASA Science Mission Directorate, the National
Science Foundation Grant No. AST-1238877, the University of Maryland,
Eotvos Lorand University (ELTE), the Los Alamos National Laboratory,
and the Gordon and Betty Moore Foundation.

The research leading to these results has received funding from the European 
Research Council under the European Union's Seventh Framework Programme 
(FP/2007-2013) / ERC Grant Agreement n. 308024. This project is supported by 
CONICYT's PCI program through grant DPI20140066. C.N. acknowledges support from 
CONICYT-PCHA grant Doctorado Nacional 2015-21151643. S.K. thanks Bernie Shiao 
for the assistance in retrieving the PS1 data. C.N. thanks Francisco Aros for 
his useful suggestion on the improvements of the figures.

% \bibliographystyle{mn2e}
%  \bibliography{biblio}

\begin{thebibliography}{45}
\expandafter\ifx\csname natexlab\endcsname\relax\def\natexlab#1{#1}\fi

\bibitem[{{Balbinot} \& {Gieles}(2017)}]{Balbinot17}
{Balbinot} E., {Gieles} M., 2017, ArXiv e-prints

\bibitem[{{Balbinot} {et~al}\mbox{.}(2016){Balbinot}, {Yanny}, {Li},
  {Santiago}, {Marshall}, {Finley}, {Pieres}, {Abbott}, {Abdalla}, {Allam},
  {Benoit-L{\'e}vy}, {Bernstein}, {Bertin}, {Brooks}, {Burke}, {Carnero
  Rosell}, {Carrasco Kind}, {Carretero}, {Cunha}, {da Costa}, {DePoy}, {Desai},
  {Diehl}, {Doel}, {Estrada}, {Flaugher}, {Frieman}, {Gerdes}, {Gruen},
  {Gruendl}, {Honscheid}, {James}, {Kuehn}, {Kuropatkin}, {Lahav}, {March},
  {Martini}, {Miquel}, {Nichol}, {Ogando}, {Romer}, {Sanchez}, {Schubnell},
  {Sevilla-Noarbe}, {Smith}, {Soares-Santos}, {Sobreira}, {Suchyta}, {Tarle},
  {Thomas}, {Tucker}, {Walker}, \& {DES Collaboration}}]{Balbinot16}
{Balbinot} E. {et~al.}, 2016, \apj, 820, 58

\bibitem[{{Belokurov} {et~al}\mbox{.}(2006){Belokurov}, {Evans}, {Irwin},
  {Hewett}, \& {Wilkinson}}]{Belokurov2006}
{Belokurov} V., {Evans} N.~W., {Irwin} M.~J., {Hewett} P.~C., {Wilkinson}
  M.~I., 2006, \apjl, 637, L29

\bibitem[{{Bernard} {et~al}\mbox{.}(2014){Bernard}, {Ferguson}, {Schlafly},
  {Abbas}, {Bell}, {Deacon}, {Martin}, {Rix}, {Sesar}, {Slater},
  {Pe{\~n}arrubia}, {Wyse}, {Burgett}, {Chambers}, {Draper}, {Hodapp},
  {Kaiser}, {Kudritzki}, {Magnier}, {Metcalfe}, {Morgan}, {Price}, {Tonry},
  {Wainscoat}, \& {Waters}}]{Bernard2014}
{Bernard} E.~J. {et~al.}, 2014, \mnras, 443, L84

\bibitem[{{Bernard} {et~al}\mbox{.}(2016){Bernard}, {Ferguson}, {Schlafly},
  {Martin}, {Rix}, {Bell}, {Finkbeiner}, {Goldman}, {Mart{\'{\i}}nez-Delgado},
  {Sesar}, {Wyse}, {Burgett}, {Chambers}, {Draper}, {Hodapp}, {Kaiser},
  {Kudritzki}, {Magnier}, {Metcalfe}, {Wainscoat}, \& {Waters}}]{Bernard2016}
{Bernard} E.~J. {et~al.}, 2016, \mnras, 463, 1759

\bibitem[{{Bonaca} {et~al}\mbox{.}(2012){Bonaca}, {Geha}, \&
  {Kallivayalil}}]{Bonaca12}
{Bonaca} A., {Geha} M., {Kallivayalil} N., 2012, \apjl, 760, L6

\bibitem[{{Bowden} {et~al}\mbox{.}(2015){Bowden}, {Belokurov}, \&
  {Evans}}]{Bowden15}
{Bowden} A., {Belokurov} V., {Evans} N.~W., 2015, \mnras, 449, 1391

\bibitem[{{Carballo-Bello} {et~al}\mbox{.}(2012){Carballo-Bello}, {Gieles},
  {Sollima}, {Koposov}, {Mart{\'{\i}}nez-Delgado}, \&
  {Pe{\~n}arrubia}}]{Julio12}
{Carballo-Bello} J.~A., {Gieles} M., {Sollima} A., {Koposov} S.,
  {Mart{\'{\i}}nez-Delgado} D., {Pe{\~n}arrubia} J., 2012, \mnras, 419, 14

\bibitem[{{Carballo-Bello} {et~al}\mbox{.}(2014){Carballo-Bello}, {Sollima},
  {Mart{\'{\i}}nez-Delgado}, {Pila-D{\'{\i}}ez}, {Leaman}, {Fliri},
  {Mu{\~n}oz}, \& {Corral-Santana}}]{Julio14}
{Carballo-Bello} J.~A., {Sollima} A., {Mart{\'{\i}}nez-Delgado} D.,
  {Pila-D{\'{\i}}ez} B., {Leaman} R., {Fliri} J., {Mu{\~n}oz} R.~R.,
  {Corral-Santana} J.~M., 2014, \mnras, 445, 2971

\bibitem[{{Chambers} {et~al}\mbox{.}(2016){Chambers}, {Magnier}, {Metcalfe},
  {Flewelling}, {Huber}, {Waters}, {Denneau}, {Draper}, {Farrow}, {Finkbeiner},
  {Holmberg}, {Koppenhoefer}, {Price}, {Saglia}, {Schlafly}, {Smartt},
  {Sweeney}, {Wainscoat}, {Burgett}, {Grav}, {Heasley}, {Hodapp}, {Jedicke},
  {Kaiser}, {Kudritzki}, {Luppino}, {Lupton}, {Monet}, {Morgan}, {Onaka},
  {Stubbs}, {Tonry}, {Banados}, {Bell}, {Bender}, {Bernard}, {Botticella},
  {Casertano}, {Chastel}, {Chen}, {Chen}, {Cole}, {Deacon}, {Frenk},
  {Fitzsimmons}, {Gezari}, {Goessl}, {Goggia}, {Goldman}, {Grebel}, {Hambly},
  {Hasinger}, {Heavens}, {Heckman}, {Henderson}, {Henning}, {Holman}, {Hopp},
  {Ip}, {Isani}, {Keyes}, {Koekemoer}, {Kotak}, {Long}, {Lucey}, {Liu},
  {Martin}, {McLean}, {Morganson}, {Murphy}, {Nieto-Santisteban}, {Norberg},
  {Peacock}, {Pier}, {Postman}, {Primak}, {Rae}, {Rest}, {Riess}, {Riffeser},
  {Rix}, {Roser}, {Schilbach}, {Schultz}, {Scolnic}, {Szalay}, {Seitz},
  {Shiao}, {Small}, {Smith}, {Soderblom}, {Taylor}, {Thakar}, {Thiel},
  {Thilker}, {Urata}, {Valenti}, {Walter}, {Watters}, {Werner}, {White},
  {Wood-Vasey}, \& {Wyse}}]{Chambers16}
{Chambers} K.~C. {et~al.}, 2016, ArXiv e-prints

\bibitem[{{Cohen} \& {Melendez}(2005)}]{Cohen2005}
{Cohen} J.~G., {Melendez} J., 2005, \aj, 129, 1607

\bibitem[{{Cote} {et~al}\mbox{.}(1991){Cote}, {Richer}, \& {Fahlman}}]{Cote91}
{Cote} P., {Richer} H.~B., {Fahlman} G.~G., 1991, \aj, 102, 1358

\bibitem[{{Dehnen} {et~al}\mbox{.}(2004){Dehnen}, {Odenkirchen}, {Grebel}, \&
  {Rix}}]{Dehnen04}
{Dehnen} W., {Odenkirchen} M., {Grebel} E.~K., {Rix} H.-W., 2004, \aj, 127,
  2753

\bibitem[{{Erkal} \& {Belokurov}(2015{\natexlab{a}})}]{Erkal15a}
{Erkal} D., {Belokurov} V., 2015{\natexlab{a}}, \mnras, 450, 1136

\bibitem[{{Erkal} \& {Belokurov}(2015{\natexlab{b}})}]{Erkal2015b}
{Erkal} D., {Belokurov} V., 2015{\natexlab{b}}, \mnras, 454, 3542

\bibitem[{{Erkal} {et~al}\mbox{.}(2016){Erkal}, {Koposov}, \&
  {Belokurov}}]{Erkal16}
{Erkal} D., {Koposov} S.~E., {Belokurov} V., 2016, ArXiv e-prints

\bibitem[{{Eyre} \& {Binney}(2011)}]{Eyre11}
{Eyre} A., {Binney} J., 2011, \mnras, 413, 1852

\bibitem[{{Farrow} {et~al}\mbox{.}(2014){Farrow}, {Cole}, {Metcalfe}, {Draper},
  {Norberg}, {Foucaud}, {Burgett}, {Chambers}, {Kaiser}, {Kudritzki},
  {Magnier}, {Price}, {Tonry}, \& {Waters}}]{Farrow2014}
{Farrow} D.~J. {et~al.}, 2014, \mnras, 437, 748

\bibitem[{{Figuera Jaimes} {et~al}\mbox{.}(2013){Figuera Jaimes}, {Arellano
  Ferro}, {Bramich}, {Giridhar}, \& {Kuppuswamy}}]{Figuera13}
{Figuera Jaimes} R., {Arellano Ferro} A., {Bramich} D.~M., {Giridhar} S.,
  {Kuppuswamy} K., 2013, \aap, 556, A20

\bibitem[{{Flewelling} {et~al}\mbox{.}(2016){Flewelling}, {Magnier},
  {Chambers}, {Heasley}, {Holmberg}, {Huber}, {Sweeney}, {Waters}, {Chen},
  {Farrow}, {Hasinger}, {Henderson}, {Long}, {Metcalfe}, {Nieto-Santisteban},
  {Norberg}, {Saglia}, {Szalay}, {Rest}, {Thakar}, {Tonry}, {Valenti},
  {Werner}, {White}, {Denneau}, {Draper}, {Hodapp}, {Jedicke}, {Kaiser},
  {Kudritzki}, {Price}, {Wainscoat}, {Chastel}, {McClean}, {Postman}, \&
  {Shiao}}]{Flewelling16}
{Flewelling} H.~A. {et~al.}, 2016, ArXiv e-prints

\bibitem[{{Gnedin} {et~al}\mbox{.}(1999){Gnedin}, {Lee}, \&
  {Ostriker}}]{Gnedin99}
{Gnedin} O.~Y., {Lee} H.~M., {Ostriker} J.~P., 1999, \apj, 522, 935

\bibitem[{{Gnedin} \& {Ostriker}(1997)}]{Gnedin97}
{Gnedin} O.~Y., {Ostriker} J.~P., 1997, \apj, 474, 223

\bibitem[{{Grillmair} \& {Dionatos}(2006{\natexlab{a}})}]{Grillmair2006}
{Grillmair} C.~J., {Dionatos} O., 2006{\natexlab{a}}, \apjl, 641, L37

\bibitem[{{Grillmair} \& {Dionatos}(2006{\natexlab{b}})}]{Grillmair06}
{Grillmair} C.~J., {Dionatos} O., 2006{\natexlab{b}}, \apjl, 643, L17

\bibitem[{{Harris}(1996)}]{Harris96}
{Harris} W.~E., 1996, \aj, 112, 1487

\bibitem[{{Koposov} {et~al}\mbox{.}(2014){Koposov}, {Irwin}, {Belokurov},
  {Gonzalez-Solares}, {Yoldas}, {Lewis}, {Metcalfe}, \& {Shanks}}]{Koposov14}
{Koposov} S.~E., {Irwin} M., {Belokurov} V., {Gonzalez-Solares} E., {Yoldas}
  A.~K., {Lewis} J., {Metcalfe} N., {Shanks} T., 2014, \mnras, 442, L85

\bibitem[{{Koposov} {et~al}\mbox{.}(2010){Koposov}, {Rix}, \&
  {Hogg}}]{Koposov10}
{Koposov} S.~E., {Rix} H.-W., {Hogg} D.~W., 2010, \apj, 712, 260

\bibitem[{{K{\"u}pper} {et~al}\mbox{.}(2015){K{\"u}pper}, {Balbinot}, {Bonaca},
  {Johnston}, {Hogg}, {Kroupa}, \& {Santiago}}]{Kupper15}
{K{\"u}pper} A.~H.~W., {Balbinot} E., {Bonaca} A., {Johnston} K.~V., {Hogg}
  D.~W., {Kroupa} P., {Santiago} B.~X., 2015, \apj, 803, 80

\bibitem[{{Lauchner} {et~al}\mbox{.}(2006){Lauchner}, {Powell}, \&
  {Wilhelm}}]{Lauchner2006}
{Lauchner} A., {Powell}, Jr. W.~L., {Wilhelm} R., 2006, \apjl, 651, L33

\bibitem[{{Law} \& {Majewski}(2010)}]{LM10}
{Law} D.~R., {Majewski} S.~R., 2010, \apj, 718, 1128

\bibitem[{{Lee} {et~al}\mbox{.}(2004){Lee}, {Lee}, {Fahlman}, \&
  {Sung}}]{Lee04}
{Lee} K.~H., {Lee} H.~M., {Fahlman} G.~G., {Sung} H., 2004, \aj, 128, 2838

\bibitem[{{Leon} {et~al}\mbox{.}(2000){Leon}, {Meylan}, \& {Combes}}]{Leon00}
{Leon} S., {Meylan} G., {Combes} F., 2000, \aap, 359, 907

\bibitem[{{Lindegren} {et~al}\mbox{.}(2016){Lindegren}, {Lammers}, {Bastian},
  {Hern{\'a}ndez}, {Klioner}, {Hobbs}, {Bombrun}, {Michalik}, {Ramos-Lerate},
  {Butkevich}, {Comoretto}, {Joliet}, {Holl}, {Hutton}, {Parsons},
  {Steidelm{\"u}ller}, {Abbas}, {Altmann}, {Andrei}, {Anton}, {Bach},
  {Barache}, {Becciani}, {Berthier}, {Bianchi}, {Biermann}, {Bouquillon},
  {Bourda}, {Br{\"u}semeister}, {Bucciarelli}, {Busonero}, {Carlucci},
  {Casta{\~n}eda}, {Charlot}, {Clotet}, {Crosta}, {Davidson}, {de Felice},
  {Drimmel}, {Fabricius}, {Fienga}, {Figueras}, {Fraile}, {Gai}, {Garralda},
  {Geyer}, {Gonz{\'a}lez-Vidal}, {Guerra}, {Hambly}, {Hauser}, {Jordan},
  {Lattanzi}, {Lenhardt}, {Liao}, {L{\"o}ffler}, {McMillan}, {Mignard}, {Mora},
  {Morbidelli}, {Portell}, {Riva}, {Sarasso}, {Serraller}, {Siddiqui}, {Smart},
  {Spagna}, {Stampa}, {Steele}, {Taris}, {Torra}, {van Reeven}, {Vecchiato},
  {Zschocke}, {de Bruijne}, {Gracia}, {Raison}, {Lister}, {Marchant},
  {Messineo}, {Soffel}, {Osorio}, {de Torres}, \& {O'Mullane}}]{Lindegren2016}
{Lindegren} L. {et~al.}, 2016, \aap, 595, A4

\bibitem[{{Magnier} {et~al}\mbox{.}(2016{\natexlab{a}}){Magnier}, {Chambers},
  {Flewelling}, {Hoblitt}, {Huber}, {Price}, {Sweeney}, {Waters}, {Denneau},
  {Draper}, {Hodapp}, {Jedicke}, {Kaiser}, {Kudritzki}, {Metcalfe}, {Stubbs},
  \& {Wainscoast}}]{Magnier2016a}
{Magnier} E.~A. {et~al.}, 2016{\natexlab{a}}, ArXiv e-prints

\bibitem[{{Magnier} {et~al}\mbox{.}(2016{\natexlab{b}}){Magnier}, {Sweeney},
  {Chambers}, {Flewelling}, {Huber}, {Price}, {Waters}, {Denneau}, {Draper},
  {Jedicke}, {Hodapp}, {Kaiser}, {Kudritzki}, {Metcalfe}, {Stubbs}, \&
  {Wainscoast}}]{Magnier2016b}
{Magnier} E.~A. {et~al.}, 2016{\natexlab{b}}, ArXiv e-prints

\bibitem[{{Michalik} {et~al}\mbox{.}(2015){Michalik}, {Lindegren}, \&
  {Hobbs}}]{Michalik2015}
{Michalik} D., {Lindegren} L., {Hobbs} D., 2015, \aap, 574, A115

\bibitem[{{Niederste-Ostholt} {et~al}\mbox{.}(2010){Niederste-Ostholt},
  {Belokurov}, {Evans}, {Koposov}, {Gieles}, \& {Irwin}}]{NO10}
{Niederste-Ostholt} M., {Belokurov} V., {Evans} N.~W., {Koposov} S., {Gieles}
  M., {Irwin} M.~J., 2010, \mnras, 408, L66

\bibitem[{{Odenkirchen} {et~al}\mbox{.}(2003){Odenkirchen}, {Grebel}, {Dehnen},
  {Rix}, {Yanny}, {Newberg}, {Rockosi}, {Mart{\'{\i}}nez-Delgado}, {Brinkmann},
  \& {Pier}}]{Odenkirchen2003}
{Odenkirchen} M. {et~al.}, 2003, \aj, 126, 2385

\bibitem[{{Odenkirchen} {et~al}\mbox{.}(2001){Odenkirchen}, {Grebel},
  {Rockosi}, {Dehnen}, {Ibata}, {Rix}, {Stolte}, {Wolf}, {Anderson}, {Bahcall},
  {Brinkmann}, {Csabai}, {Hennessy}, {Hindsley}, {Ivezi{\'c}}, {Lupton},
  {Munn}, {Pier}, {Stoughton}, \& {York}}]{Odenkirchen2001}
{Odenkirchen} M. {et~al.}, 2001, \apjl, 548, L165

\bibitem[{{Rockosi} {et~al}\mbox{.}(2002){Rockosi}, {Odenkirchen}, {Grebel},
  {Dehnen}, {Cudworth}, {Gunn}, {York}, {Brinkmann}, {Hennessy}, \&
  {Ivezi{\'c}}}]{Rockosi2002}
{Rockosi} C.~M. {et~al.}, 2002, \aj, 124, 349

\bibitem[{{Schlafly} \& {Finkbeiner}(2011)}]{Schlafly11}
{Schlafly} E.~F., {Finkbeiner} D.~P., 2011, \apj, 737, 103

\bibitem[{{Schlegel} {et~al}\mbox{.}(1998){Schlegel}, {Finkbeiner}, \&
  {Davis}}]{SFD}
{Schlegel} D.~J., {Finkbeiner} D.~P., {Davis} M., 1998, \apj, 500, 525

\bibitem[{{Sollima} {et~al}\mbox{.}(2011){Sollima}, {Mart{\'{\i}}nez-Delgado},
  {Valls-Gabaud}, \& {Pe{\~n}arrubia}}]{Solima11}
{Sollima} A., {Mart{\'{\i}}nez-Delgado} D., {Valls-Gabaud} D., {Pe{\~n}arrubia}
  J., 2011, \apj, 726, 47

\bibitem[{{Yoon} {et~al}\mbox{.}(2011){Yoon}, {Johnston}, \& {Hogg}}]{Yoon11}
{Yoon} J.~H., {Johnston} K.~V., {Hogg} D.~W., 2011, \apj, 731, 58

\end{thebibliography}

%\listofchanges
\end{document}